\documentclass[12pt]{article}
\usepackage[superscript]{cite} 
\usepackage{hyperref}
\usepackage{times}
\usepackage[T1]{fontenc} 
\usepackage{breakcites} 
\usepackage{url} 

\newcommand{\itemb}{\begin{itemize}}
\newcommand{\iteme}{\end{itemize}}

\newcommand{\enumb}{\begin{enumerate}}
\newcommand{\enume}{\end{enumerate}}

\newcommand{\smallest}[1]{\vspace{6pt}\noindent {\bf  {#1}.}}

\newcommand{\com}[1]{}
\usepackage{xcolor}

\newcommand{\brr}{baroreceptor}

\newcommand{\nep}{norepinephrine}

\newcommand{\indep}{independent}

\newcommand{\xtblty}{excitability}

\newcommand{\hypog}{hypoglycemia}

\newcommand{\desensit}{desensitization}
\newcommand{\Desensit}{Desensitization}

\newcommand{\atn}{{{autonomic nervous system}}}

\newcommand{\sympa}{{{sympathetic}}}
\newcommand{\parasympa}{{{parasympathetic}}}

\newcommand{\sns}{{{sympathetic nervous system}}}

\newcommand{\SNS}{{{Sympathetic Nervous System}}}

\newcommand{\vasodil}{{{vasodilation}}}

\newcommand{\vasoconst}{{{vasoconstriction}}}

\newcommand{\homeos}{{{homeostasis}}}

\newcommand{\catwo}{{{Ca}\textsuperscript{{\raisebox{-1pt}{$\scriptstyle{2+}$}}}}}

\newcommand{\nktpase}{{{Na}\textsuperscript{+}/{K}\textsuperscript{+}-ATPase}}

\newcommand{\thn}{{T*MGR}} %

\newenvironment{sciabstract}{\begin{quote} \bf}{\end{quote}}

\usepackage{fancyhdr}
\title{A \SNS\ \\Theory of Migraine}
\author{
{Ari Rappoport}\\ 
\normalsize{The Hebrew University of Jerusalem, Israel}\\
\normalsize{ari.rappoport@mail.huji.ac.il}
}
\date{\normalsize{July 2024}}
\begin{document}
\maketitle
\begin{sciabstract}
Migraine (MGR) ranks first among diseases in terms of years of lost healthy life in young adult and adult women. Currently, there is no theory of MGR. 
This paper presents a complete theory of migraine that explains its etiology, symptoms, pathology, and treatment. 
Migraine involves partially saturated (usually chronically high) \sns\ (SNS) activity, mainly due to higher sensitivity of the metabolic sensors that recruit it. MGR headache occurs when SNS activity is desensitized or excessive, resulting in hyper\xtblty\ of baroreceptors, oxidative stress, and activation of pain pathways via TRPV1 channels and CGRP. The theory is supported by overwhelming evidence, and explains the properties of current MGR treatments. 
\end{sciabstract}

\section{Introduction}
Migraine (MGR) ranks second among diseases in terms of years of lost healthy life worldwide, and the first among women between the ages 15 and 49 \cite{steiner2020mig}. 
As such, MGR is one of the most important human diseases, and the most common brain-associated disease. 
Although much progress has been made in recent decades in the understanding of MGR and in new drug treatment, both the cause of the disease and its mechanisms are still largely unknown \cite{ferrari2022mig, ashina2020mig, haanes2019pat, goadsby2017pat}. 

This paper presents the first complete theory of migraine (\thn). It is complete in the sense that it mechanistically explains the etiology, symptoms, pathophysiology, and treatment of the disease. \thn\ is supported by overwhelming evidence, and addresses all of the salient empirical data known about the disease. 

\smallest{Theory overview}
Briefly, \thn\ explains MGR using four simple steps. First, at the basal state, MGR involves partial saturation of the \nep\ (NEP) branch of the \sns\ (SNS). This can be due to chronically increased activity, or to inherently reduced capacity. The problem can stem from many causes, the most supported cause in MGR being higher sensitivity of the sensors triggering SNS activation. 

Second, in the acute phase, MGR headache is caused by \desensit\ and/or excessive activation of the SNS, due to its partial saturation and to various triggers that increase SNS activity. 

Third, SNS \desensit\ affects vascular smooth muscle cells (VSMCs) and \brr s (BRRs), which are continuously active to regulate blood pressure (BPR). These cells rely on the NEP SNS for their phasic function, which reduces BPR via the \parasympa\ nervous system (PNS). Impaired NEP SNS function reduces their available energy, specifically impairing their \nktpase. This leads to calcium entry and hyper\xtblty. Even without \desensit, excessive SNS activation can induce hyper\xtblty. 

Finally, the increased BRR firing and increased VSMC contractility yield oxidative and nitrosative stress, which activates TRPV1 channels. These in turn induce the release of the neuropeptide CGRP. Normally, these agents reduce BRR activation to allow for higher BPR during effort. However, hyper\xtblty\ interferes with this normal function, causing prolonged activation instead. Both TRPV1 and CGRP are strongly expressed in pain pathways, and convey pain signals to the brain. This is MGR headache.  

A non-MGR contribution of \thn\ is a new physiological role for CGRP, whereby it mediates adaptation to higher stress levels. Specifically to BRRs, it mediates the phenomenon known as BRR resetting. 

All four parts of this account are supported by strong bodies of evidence related to general and MGR patient data about the SNS, the vasculature, BRRs, CGRP, TRPV1, MGR triggers and premonitory symptoms, and agents used for MGR treatment. 

\section{Theory}
\subsection{SNS and BRRs}
The \atn, comprised of the SNS and the PNS, is activated in states of deviation from \homeos\ in order to help restore it \cite{benarroch2007aut}. The SNS is activated during deficiency states, including \hypog\ (hypog), hypoxia, cold, pain, and psychological stress (a deficiency state because it increases the metabolic demands from the body). In these states, the SNS increases heart rate and blood pressure, increases blood glucose via lipolysis and glycogenolysis, induces thermogenesis, and induces \vasodil\ in performance tissues (the brain, skeletal muscle, cardiovasculature) \cite{benarroch2007aut}. 
The SNS uses both NEP and epinephrine (EPI). 

A few lesser known facts about the SNS are highly relevant for MGR. First, 
SNS control over \brr s is done mainly via NEP rather than EPI \cite{morrison2001dif}. 
Second, although sweating is known to be an exception in that it is an SNS function mediated by post-ganglionic acetylcholine (ACh) rather than NEP or EPI, NEP can in fact induce sweating \cite{chemali2001alp}, 
with heat-induced skin \vasodil\ partially mediated by the SNS \cite{hodges2009inv}. 
Indeed, the superior cervical ganglion provides SNS innervation of lacrimal glands, in addition to BRRs, cranial muscle blood vessels, pupil dilators, choroid plexus, and the pineal gland. 
Third, the SNS not only increases blood glucose levels, it also stimulates glucose uptake by cells via insulin-\indep\ translocation of the Glut4 transporter to the plasma membrane \cite{dehvari2012bet}. 

\smallest{\Desensit}
SNS responses desensitize after some time, including the responses  to hypog \cite{kakall2019rep}, 
cold \cite{cernecka2014odd}, 
and hypoxia \cite{long2002fet} (for BRRs, see below). 
\Desensit\ involves endocytosis and degradation of the NEP receptors \cite{okeke2019ago}, but can also involve vesicle depletion. Agonist-induced \desensit\ of beta3-NEP occurs much more slowly than that of beta1-NEP, beta2-NEP \cite{okeke2019ago}, and alpha1-NEP \cite{maze1985epi} (days vs.\ minutes to hours).  


\smallest{Baroreceptors}
BRRs are sensory neurons mainly located at the carotid bodies and heart aorta \cite{hall2020guy}. 
They are stretch-sensitive mechanoreceptors, firing when BPR increases. During non-stress conditions, they initiate the baroreflex, which maintains a constant BPR level via negative feedback mechanisms. BRRs directly excite neurons in the nucleus of the solitary tract (NTS), which excite GABAergic neurons in the caudal ventrolateral medulla (VLM), which in turn suppress rostral VLM (RVLM) neurons exciting \sympa\ preganglionic neurons that control the heart and blood vessels \cite{schreihofer2002bar}. 
Thus, the net effect of the baroreflex is to suppress the SNS and allow a parasympathetic tone, which decreases BPR and heart rate.  
Decreases in BPR allow SNS responses, to start a new BRR cycle. 

Under effortful conditions where BPR is increased for a prolonged period of time (e.g., exercise) \cite{liu2021sin}, 
BRRs undergo `resetting' in which their activation only occurs at the new, higher BPR \cite{wehrwein2013reg}. 
The precise mechanisms behind resetting are not known, but it is known to involve direct damage to mechanosensitive BRRs, loss of BRR coupling to vascular wall stretch, and increased blood vessel stiffness \cite{wehrwein2013reg, dampney2017res}.

Crucially for MGR, VSMCs completely rely on lactate to provide ATP for the activation of the \nktpase\ \cite{shi2020met}. 
In general, lactate can be produced by aerobic glycolysis and glycogenolysis, or be shuttled to neurons from astrocytes. All of these processes, and the \nktpase, are stimulated by NEP, in the brain and skeletal muscle \cite{clausen2000rol, gibbs2008rol, jensen2020inh}, and in BRRs \cite{akoev1980cat}. 
Indeed, there is rich SNS innervation of carotid and aortic BRRs \cite{akoev1980cat}, 
BRRs contain dense glycogen granules and are surrounded by cells resembling Schwann cells (the peripheral astrocytes) \cite{tranum1975ult}, 
and suppression of VSM \nktpase\ induces hypertension \cite{sekine1984na}. 

Chemoreceptors, which sense oxygen and CO2, are located in close proximity to BRRs, and also have a role in MGR, although it is smaller than that of BRRs (see below). 

\subsection{TRPV1, CGRP}
\smallest{TRPV1}
TRPV1 is a member of the family of TRP channels, which are expressed in sensory neurons and sense heat, cold, stretch, oxidative stress, etc.\ \cite{nilius2014tra}. 
In particular, BRRs express TRPV1 \cite{sun2009sen}, and it is needed for their proper functioning \cite{sun2009sen}. 
Another TRP family member, TRPA1, is expressed in chemoreceptors \cite{wang2010trp}. 

TRPV1 is usually described as a heat sensor. However, it is also an oxidative stress sensor \cite{starr2008rea, nishio2013rea, chen2024rea, ibi2008rea, negri2022con, kievit2022mit}. 
TRPV1 is well-known as a conveyor of pain \cite{nilius2014tra}, including when it is activated by oxidative stress \cite{ibi2008rea, nishio2013rea, chen2024rea}. 

TRP channels are quickly desensitized by agonists, but TRPV1 is unique in that it can be reactivated after complete \desensit\ by higher agonist concentrations \cite{novakova2007fun}. 

\smallest{CGRP}
CGRP is a neuropeptide mainly expressed in sensory neurons, including in perivascular fibers, lungs, gut, immune cells, spinal cord, and trigeminal ganglia \cite{russell2014cal, russo2023cgrp}. In cortex, it is expressed in pain areas \cite{russo2023cgrp}. 
Its release is stimulated by neural depolarization, including by TRPV1 in trigeminal sensory neurons \cite{akerman2003van}. 
Its best-known function is that of being a highly potent vasodilator (via KATP channels), and as a result, its role is usually described as being protective \cite{russell2014cal, russo2023cgrp}. 
It is clearly involved in MGR (see below). 

Here I introduce a new account of CGRP, where its role is to re-calibrate activation thresholds along its path in light of prolonged stress. Because the focus of the present paper is MGR, I will focus on evidence that CGRP has this role in the cardiovascular system. 

The hypothesis relevant here is that {\bf CGRP is a major mediator of BRR resetting}. As such, it reduces SNS activation during prolonged effort/stress. Indeed, CGRP release is significantly increased during exercise and heart infarction \cite{lechleitner1994exe}. 
It recovers skeletal muscle \xtblty\ during muscle contraction by stimulating \nktpase\ activation, and maintains force development \cite{clausen2000rol, macdonald2008eff}. 
Importantly, CGRP knockout yields higher SNS activation (higher BPR, heart rate, BRR sensitivity, urine NEP/EPI metabolites) \cite{oh2001ele, mai2014cal}. 
CGRP in the lower brainstem decreases BRR sensitivity \cite{kim1998end}. 
CGRP opposes SNS-induced thermogenesis \cite{osaka1998tem}, 
and non-effortful SNS activation opposes CGRP release \cite{bowles2003bet, marichal2020mon}
while NEP reuptake inhibitors (simulating a high SNS activation state) increase CGRP in adrenal cathecholamine neurons \cite{hofle1991sti}. 


\subsection{Migraine}
\thn\ is a simple theory of MGR. MGR involves an altered basal state in which the SNS is partially saturated, due to chronically increased activity or inherently reduced capacity. Headache attacks occur in two scenarios. First, they occur when triggers lead to \desensit\ of BRRs or of the whole NEP branch of the SNS. Because BRRs are the only place where NEP SNS is continuously used, it is also the first place to suffer the consequences of \desensit. Since VSMCs and BRRs rely on NEP to provide energy for the \nktpase, NEP \desensit\ leads to \nktpase\ failure and hence cellular hyper\xtblty. This induces \catwo\ entry via voltage-gated calcium channels, resulting in excitotoxicity and oxidative stress \cite{connolly2015met}. 
Oxidative stress activates TRPV1 channels, which induce CGRP release. Both of these agents excite downstream brain pain pathways. 

A second scenario in which headache can occur is via excessive activation of the SNS (due to saturation and triggers) without its \desensit. This can also lead to BRR hyperactivity and pain as described in the first scenario. 

Here we will see that this account explains the properties of MGR (detailed evidence is given in the next section). 

\smallest{Triggers}
Basically all of the conditions known to trigger MGR or claimed to  do so \cite{kelman2007tri, gross2019met} increase SNS activity and/or TRPV1/CGRP.  
Exercise, skipping meals (hypog), stress, hypoxia and smoke directly stimulate the SNS. 
Dehydration increases BRR activity \cite{charkoudian2003inf}. 
Heat initially decreases skin SNS activity, but this is followed by a strong increase with sweating \cite{gagnon2018swe}. 
Menstruation involves higher body temperature \cite{baker2020tem}, so the effect is similar to heat. 
Alcohol induces hypog and exacerbates insulin-induced hypog \cite{oba2021com}. 
Alcohol also directly recruits TRPV1 and CGRP \cite{nicoletti2008eth}. 

Glyceryl trinitrate (GTN, nitroglycerin) has been shown to induce headache with MGR characteristics in patients \cite{schoonman2008mig}. 
GTN, and other nitric oxide donors, directly increase CGRP release (4x in rat trigeminal ganglia) and CGRP promoter activity \cite{bellamy2006nit}. 
The induction of MGR attacks by vasodilators (GTN, KATP openers) \cite{alKaragholi2019ope} can also be explained via strong stimulation of BRRs. 

\smallest{Premonitory symptoms}
MGR attacks are commonly preceded by so-called premonitory symptoms. All of these symptoms indicate either higher SNS activity or its ongoing \desensit. 

Thirst is a very common premonitory symptom \cite{giffin2003pre}. It is induced by many stimuli that induce renin release, including the SNS, carotid artery \vasoconst\ (BRRs), and NEP/EPI \cite{fitzsimons1975ren}. 
Hunger and food craving are due to activation of the counter-regulatory responses to hypog, which include the SNS \cite{kakall2019rep}. 
Nausea and vomiting are known symptoms of the postural tachycardia syndrome (POTS), a dysautonomia that involves SNS hyperactivation \cite{shibao2005hyp}. 
Urination \cite{giffin2003pre} is an SNS-mediated response. 
Tiredness and concentration difficulties \cite{giffin2003pre} indicate that SNS \desensit\ is underway. 
Neck pain or stiffness occur in more than half of MGR patients \cite{blau1994mig, giffin2003pre}, showing that this location is central to MGR headache. 

\smallest{Nature of pain}
MGR headache is unique among headaches in that it is described as `throbbing' (pulsing). This happens because it is generated by repeated BRR cellular firing. It has been argued \cite{goadsby2017pat} that since the throbbing pain rate is not synchronized with the arterial pulse rate \cite{ahn2010tem}, the vasculature is not involved in MGR pain. However, MGR pain involves cellular hyper\xtblty, so the rate of pain transmission is indeed expected to be different from that of heart beats. 

\smallest{Timeline of pain}
MGR headache can build up over days, and  takes  hours to days to pass. This occurs when the problem and/or the triggers focus on the beta3-NEP receptor (supporting thermogenesis), which requires this time frame to desensitize and resensitize \cite{okeke2019ago}. 

\smallest{Aura}
A subset of MGR patients experience aura symptoms during (and before) attacks, mainly including flashes of light and blind spots. 
\thn\ explains aura as follows. The SNS affects visual inputs to the brain via beta3-NEP receptors expressed in retinal blood vessels \cite{mori2010pha}. 
SNS \desensit\ disrupts normal visual processing, inducing hyperactivity as in BRRs. Indeed, human visual cortex shows elevated activity during aura \cite{hadjikhani2001mec}. 
Aura is a special case of the next item. 

\smallest{Sensory hypersensitivity}
Vision is not the only modality that shows sensory hypersensitivity in MGR. In fact, the idea that MGR is basically a sensory hypersensitivity disease has been raised \cite{goadsby2017pat}. 
Virtually all low-level sensory pathways are affected by the SNS and express TRPV1 and CGRP. 

\section{Evidence}
To support \thn, we need evidence from MGR patients showing partial SNS saturation, why this might happen, SNS \desensit\ or excessive activation right before and during attacks, BRR dysfunction, and high oxidative stress, TRPV1, and CGRP in the MGR paths. 
All of this evidence already exists in the literature and is listed below. 

\subsection{Basal state}
\smallest{Partial SNS saturation (SNS hypofunction)}
A large number of papers have used a battery of tests to assess SNS and PNS function at the pain-free (inter-ictal) period , including SNS/PNS responses to quick standing, the tilt test, deep breathing, handgrip, the Valsalva maneuver, thermal (cold pain) sensitivity, psychological stress, skin sweating (SCR), heart rate variability, and hyperCO2.  The vast majority (but not all \cite{pierangeli1997pow})  
of these papers report SNS hypofunction \cite{cohen1983psy, drummond1985vas,  rubin1985aut, cortelli1986car, havanka1986aut, havanka1987car, gomi1989swe, tabata2000cos, harer1991cer, martin1992car, pogacnik1993aut,  fiermonte1995cer, mosek1999aut, vernieri2008inc, yildiz2008sym, venkatesan2014stu, cambron2014aut, mamontov2016aut,  koenig2016vag, miglis2018mig, zhang2021hea, kavuncu2022shi, shi2022fun, pavelic2024rec}. 

This indicates that the SNS is close to saturation. The most likely reasons are that it is already active at a higher capacity than normal, so that its capacity to respond to rapid challenges is decreased, or that it is already partially desensitized. However, an inherent hypofunction, supported by some evidence \cite{gotoh1984nor, mikamo1989car}, is also possible. 


The data above can also be interpreted to show higher PNS activity, which dominates the SNS. However, the PNS generally also shows hypofunction in MGR \cite{thomsen1995tra} (also shown in some of the tests cited above), which accords with SNS hypofunction because PNS activity is largely driven by SNS reflexes. 

MGR is comorbid with several diseases associated with impaired SNS function, including respiratory diseases (asthma and chronic obstructive pulmonary disease) \cite{davey2002ass, kang2021ass}, 
anxiety, obesity, sleep problems \cite{caponnetto2021com}, 
coronary heart disease \cite{kalkman2023mig}, 
angina pectoris, myocardial infarction \cite{rose2004mig}, 
syncope \cite{khurana2018syn}, 
and POTS \cite{mueller2022pos}. 

MGR patients show brain glucose deficiency, mainly in pain areas, supporting impaired SNS/NEP function \cite{kim2010int}. 
When waking up with a headache, patients show increased NEP and cortisol 3 hours before waking \cite{hsu1977ear}. This supports hypog, since NEP and cortisol are part of the response to hypog. 
Patients also show brain ATP depletion \cite{paemeleire201331p}, which can be due to hypog or to hyper\xtblty. 

Small fiber neuropathy has been demonstrated in patient skin, showing peripheral autonomic damage \cite{stillman2021aut}. 

\smallest{BRR, BPR}
In addition to the general SNS saturation noted above, there is also specific evidence for altered BRR sensitivity in MGR \cite{rossi2022imp, mueller2023red}, 
and for SNS neck hypofunction \cite{korkmaz2015sym}. 
In addition, decreased BRR sensitivity predicts chronic headache \cite{mueller2023red}. 
BRR-mediated BPR responses are reduced in MGR \cite{sanya2005imp}. 
Patients show higher resting diastolic BPR \cite{shechter2002mig}. 
Following NEP injection, patients take 5x as long until BPR recovery vs.\ controls, showing BRR dysfunction \cite{gotoh1984nor}. 
Reduced NEP in BRRs has been demonstrated via exaggerated BPR responses to an alpha1-NEP agonist \cite{boccuni1989pre}. 

All drugs that lower BPR decrease headache frequency \cite{law2005hea, carcel2023eff}. 
More than 75\% of patients show comorbidities, hypertension being one of the most common ones \cite{caponnetto2021com}. Other common ones include other conditions associated with chronic SNS, such as anxiety disorders and back pain. 

Two large epidemiologic studies found an association between increased systolic BPR and *reduced* MGR (and other headache) risk \cite{tronvik2008hig}. 
The apparent contradiction with the other evidence can be explained as follows. On one hand, high basal SNS activity induces high BPR as explained above. On the other hand, low BPR may indicate that SNS \desensit\ is already taking place, in which case high BPR can be viewed as protective. We expect higher BPR at periods far from headache attacks (unless the SNS is chronically desensitized), and higher or lower BPR at periods closer to the attacks. BPR alone is not a reliable biomarker for MGR. 

Cerebral blood flow (CBF) is another unreliable marker for MGR. The SNS response under effort/stress includes increased heart rate and \vasodil, which increase CBF. Thus, as in BPR, basal CBF or CBF reactivity can be higher far from attacks. However, if the SNS is already desensitized, patients would show low basal CBF, and they might show high or low CBF close to attacks. The evidence is indeed inconclusive, showing all of these states \cite{micieli1995inc, weiller1995bra, deBenedittis1999cbf, hansen2019cer}.

\smallest{Hyper\xtblty}
As mentioned above, MGR patients show sensory hyper\xtblty. For example, visually and auditory evoked potentials show reduced habituation \cite{judit2000hab}, 
photophobia thresholds are significantly reduced, which correlates with blunted autonomic responses \cite{cortez2017alt},
and there is increased visual cortex and plasma lactate \cite{watanabe1996ele, aczel2021ide}. 

\smallest{Oxidative and nitrosative stress}
MGR patients show increased oxidative and nitrosative stress. 
Lipid peroxidation metabolites and nitric acid are increased in patient urine \cite{ciancarelli2003uri}, 
and patients with high frequency MGR have low serum anti-oxidants and high lipid peroxidation \cite{gross2021mit}. 

\smallest{Genetics}
According to \thn, main the mechanistic account of MGR is a saturated SNS. However, a complete account of the etiology of the disease needs to explain why this happens. There is some genetic evidence pointing to the sensors that trigger SNS activation. 

The strongest genes identified in GWAS are those encoding TRPM8, PRDM16, and TASK2 \cite{pietrobon2018ion}. 
TRPM8 is a cold sensor, with a SNP that reduces its expression and attenuates cold pain sensation being protective against MGR \cite{gavva2019red}. 
This shows that reduced recruitment of the SNS reduces MGR risk. 

MGR is associated with a SNP in the PRDM16 gene \cite{siokas2022dec}, which promotes development of brown adipose tissue (BAT). 
BAT supports thermogenesis and is an important SNS target tissue. 
TASK2 (encoded by KCNK5) is an acid sensor that participates in the hyperCO2 responses. 

Evidence from a large twin study points to the ATP1A2 gene, which encodes subunit 2 of the \nktpase\ \cite{nyholt2005gen}. 
However, there is also a negative result for this gene \cite{netzer2006hap}. 

Living in high altitudes increases the hypoxia SNS responses and is associated with higher MGR risk \cite{linde2017mig}. 
It is known that the human genome shows adaptations to high altitude \cite{bigham2014hum}, but this has not been shown specifically in MGR. 

A significant MGR association was found with 5 insulin receptor SNPs, which may stimulate compensating SNS hyperactivity \cite{mccarthy2001sin}. 
SNPs were also found in the glutathione anti-oxidant pathway \cite{kusumi2003glu}. 

\subsection{Acute headache state}
Attacks are induced by \desensit\ of BRRs and possibly of other locations of SNS function, or by excessive SNS activation. Thus, the evidence should mainly show BRR hypofunction, but it can also show hyperactivity, and activity in other parts of the SNS can be increased or decreased. Indeed, most reports show SNS \desensit\ during and after attacks \cite{lauritzen1983cha, cortelli1986car, drummond1990dis, vanHilten1991pla, friberg1991mig, mylius2003dys, yildiz2008sym, denuelle2008pos, bugdayci2010sal, korkmaz2015sym, miglis2018mig}.

The reduced visual and auditory habituation normalizes just before attacks, showing \desensit\ \cite{judit2000hab}. 
Some papers report high SNS activity right before and during attacks, both non-cardiovascular \cite{shaw1977met, burstein2000ass, obermann2007pre, zhang2020pla}
and cardiovascular \cite{cirignotta1982sys, perciaccante2007mig, asghar2011evi, frank2020mig}. 

The pons is activated during attacks \cite{matharu2004cen, afridi2005pos, afridi2005pet}, supporting the involvement of BRR targets. 
There is increased CGRP release \cite{goadsby1990vas}
and nitrovasive and oxidative stress during attacks \cite{yilmaz2007inc}. 
Carotid arteriography induced attacks in 9/13 patients \cite{lauritzen1983cha}, showing that the BRR location is a highly sensitive pain spot. 
In addition, the neck shows pain during attacks \cite{tfelt1981pre}. 

\smallest{Pain relievers}
Many medications for treating acute headache, including ergots and triptans, mimic the effects of the SNS (e.g., \vasoconst) without relying on it \cite{peroutka2004mig}. 
This supports SNS \desensit\ as a causative pain factor. 
Magnesium, which decreases cellular calcium influx, can also relieve pain \cite{grober2015mag}. 

\section{Treatment}
According to \thn, enhancing the SNS should be preventive in MGR. Indeed, exercise has a positive effect on the frequency and severity of headaches \cite{machado2020eff},  
and beta blockers, which oppose SNS \desensit, are probably the most effective preventive drugs \cite{ferrari2022mig}. 

Botox inhibits TRPV1 and CGRP, thereby opposing the development of pain \cite{zhang2016ext}. 

TRPV1 agonists quickly desensitize TRPV1. Indeed, intranasal civamide (TRPV1 agonist) strongly helps acute pain \cite{diamond2000int},  
and the classical TRPV1 agonist capsaicin helps in chronic MGR \cite{fusco2003rep}. 
As noted above, many sympatomimetics relieve acute pain. 

Most recently, CGRP-based drugs have been approved for treatment. 
Receptor antagonists (gepants) are moderately effective for prevention but not for acute treatment \cite{altamura2022gep}. 
This is not surprising, given that TRPV1 is expressed along all of the pain path and can convey pain even without CGRP. Moreover, CGRP calibrates the SNS and BRRs to higher workloads, and as such it should be protective in MGR in the long-term. Indeed, prolonged usage of CGRP inhibitors (gepants or CGRP antibodies) may be associated with cardiovascular risk \cite{rivera2020cgr}. 

Chocolate is claimed by some to be helpful, and by others to be a trigger \cite{gross2019met}. It is a trigger because it may yield insulin-induced hypog, and it may be acutely helpful by providing BRRs and other cells with glucose. 
However, sugars are generally not expected to provide quick effective acute treatment, because it takes time for the desensitized NEP receptors to recuperate. 

One paper reported effective acute treatment via a partial rebreathing device that increases brain oxygen, opposing hypoxia \cite{fuglsang2018tre}. 



An acute treatment direction that has not been attempted yet is lactate infusions. This sounds paradoxical in light of the basal increased lactate levels in patients (see above). However, lactate is increased at the basal state due to chronic SNS activation, while infusions during attacks may provide BRRs with the energy they need for proper \nktpase\ activation. 

\section{Discussion}
This paper presented the first complete theory of MGR. \thn\ is very simple, relying on established biological processes, and all of its components are supported by large bodies of evidence. Moreover, \thn\ addresses all of the supported evidence gathered about MGR. 

\smallest{Other theories}
There is currently no complete theory of MGR, and it is widely agreed that the cause and mechanisms of MGR are unknown \cite{ferrari2022mig, ashina2020mig, haanes2019pat, goadsby2017pat}. 
Over the years, there have been several ideas with growing and then diminishing popularity, including cortical spreading depression (a transient wave of cortical neural depolarization) and \vasodil. 
A central question in past theories was whether the pain origin is central or peripheral. The former has been convincingly ruled out \cite{olesen2009ori}, in favor of intra- and extra-cranial perivascular origin (so with some overlap with \thn). However, again,  detailed mechanisms have not been presented. 

The two theories closest to \thn\ focus on NEP and oxidative stress. MGR has been hypothesized to involve chronic SNS dysfunction with relative NEP depletion and increased release of cotransmitters dopamine, prostaglandins, ATP, and adenosine \cite{peroutka2004mig}. 
This theory overlaps \thn\ with respect to chronic SNS dysfunction, but without any mechanistic overlap or insight. 
A second theory presented MGR as a disease of brain energy deficiency that yields oxidative stress \cite{borkum2021bra}. Again, there is some shallow overlap with \thn, without detailed mechanisms. 

A great achievement of MGR research is the identification of the importance of CGRP, which has resulted in several approved CGRP-based drugs. However, as discussed above, it is not clear that such drugs are truly helpful.

\smallest{Other headaches}
It is not difficult to extend \thn\ to explain other types of headaches in addition to MGR. This was not done here in order to keep the paper focused. Nonetheless, a few words are in order. First, \thn\ trivially explains tension-type headaches, which are different from MGR by being bilateral and with a pressing rather than throbbing pain \cite{onan2023deb}. The pain mechanisms in \thn\ can be evoked without the BRR \desensit\ that lends MGR its throbbing pain aspect. 

Second, \thn\ includes a sibling theory explaining cluster headache. Cluster headache prefers men (vs.\ MGR, which prefers women), and it always involves cranial autonomic symptoms (vs.\ MGR, where these occur in many but not all patients). The main distinction between the diseases (except pain strength, which is higher in cluster headache) is probably the duration of the pain, since cluster headache attacks last between 15 and 180 minutes, and those of MGR 4-72 hours \cite{al2022deb}. 
According to \thn, cluster headache parallels MGR in that it involves the chemoreceptor paths (alpha1-NEP) rather than baroreceptors. This theory will hopefully be detailed elsewhere. 

\section*{List of Abbreviations}
{ -}

alpha1-NEP: alpha1 adrenergic receptor. 

BAT: brown adipose tissue. 

beta1/2/3-NEP: beta1/2/3/ adrenergic receptor. 

BPR: blood pressure. 

BRR: baroreptor. 

CSD: Cortical spreading depression. 

GWAS: genome-wide association studies. 

MGR: migraine. 

NEP: norepinephrine (noradrenaline). 

NTS: nucleus of the solitary tract. 

PNS: parasympatheric nervous system. 

KATP channels: ATP-inactivated potassium channels. 

POTS: postural tachycardia syndrome. 

RVLM: rostral ventrolateral medulla. 

SCR: skin conductance response. 

SNP: single-nucleotide polymorphism. 

SNS: \sns. 

TRPA1: transient receptor potential cation channel subfamily A member 1.

TRPM8: transient receptor potential cation channel subfamily M member 8.

TRPV1: transient receptor potential cation channel subfamily V member 1. 

VLM: ventrolateral medulla. 

VSMC: vascular smooth muscle cell. 

\bibliography{mgr-cluster,msc,adhd}

\begin{thebibliography}{100}
\bibitem{miglis2018mig}
Miglis MG.
\newblock Migraine and autonomic dysfunction: which is the horse and which is the jockey?
\newblock Current pain and headache reports. 2018;22:1-7.
\end{thebibliography}

\bibliographystyle{vancouver} 
\end{document}